\documentclass[aps,prd]{revtex4-2}
\usepackage[utf8]{inputenc}
\usepackage{amsmath, bm}
\usepackage{amsfonts}
\usepackage{amsthm}
\usepackage{graphicx}
\usepackage{hyperref}
\usepackage{slashed}
\usepackage{natbib}

\DeclareMathOperator{\Tr}{Tr}

\begin{document}

\title{Singularities in the RG flow}

\author{Antal Jakovác}
\email{jakovac.antal@wigner.hun-ren.hu}
\affiliation{Department of Computational Sciences, HUN-REN Wigner Research Centre for Physics, Budapest, Hungary}
\affiliation{Department of Statistics, Institute of Data Analytics and Information Systems, Corvinus University of Budapest, Hungary}

\date{\today}

%
%
%
%

\begin{abstract}
    In this work, we show examples when a perturbatively irrelevant operator becomes relevant in the infrared because of the presence of an IR singularity (IR Landau pole). An example of this behavior is the four-fermion interaction that allows the formation of bound states. The reason of the appearance of the IR Landau pole is not the singular loop as in the purely perturbative case, but the infinite number of modes appearing in the RG flow.
\end{abstract}

\maketitle

\section{Introduction}

The predictive power of a theory depends on the number of parameters it contains. The fewer the parameters, the more easy it is to tune the theory to reality, and when the parameters are determined by some observations, we may have a prediction for all other measurements.

As the branches of physics are concerned, mechanics, thermodynamics, electrodynamics, the historical success fields of physics, belong to this category. This particular simplicity allowed the construction of heat engine powered machines, fueling the Industrial Revolution in the XVIII-XIX. century.

Based on the success of these theories, simplicity became a kind of requirement for a proper physical theory. “With four parameters I can fit an elephant and with five I can make him wiggle his trunk.” said John von Neumann, signaling the deep belief of physicists in the XX. century that physics (and ultimately the world) is simple and powerful.

In field theories, including quantum field theory (QFT), the formal expression of the Lagrangian (or Hamiltonian) determines the model. The Lagrangian is represented in an operator basis, the basis coefficients are called coupling constants. According to the above expectations, a fundamental model should consist of a few basis operators. But, beyond the belief based on the historical experiences, the number of terms in a Lagrangian is not constrained in principle, and it is far from being evident why we should restrict ourselves to the simplest few.

Nevertheless, as it turned out, the Standard Model (SM) with just a small number of terms could explain all the necessary experiments performed at its relevant energy scale. It is so accurate that we are tempted to say that SM provides a full description of the observable world (maybe apart from gravity related ones, although there are a lot of efforts for their reconciliation, in renormalization group context see \cite{Nagy_2014}).

Unlike in classical theories, where the simplicity of the models often appears almost miraculous, in the context of relativistic quantum field theory (QFT), there exists a compelling rationale for omitting higher-power terms in the Lagrangian. We will delve into the details later, but here we outline a qualitative argument based on naive power counting (c.f. also \cite{Georgi1993, Kaplan2005}).

Fields carry energy dimension, while the Lagrangian must have a fixed dimension (four, in natural units). Consequently, interaction terms involving higher powers of fields must come with dimensionful coupling constants. When computing observables of fixed dimension, only dimensionless couplings can appear with arbitrary powers. As a result, couplings with negative mass dimension become suppressed at low energies: they appear multiplied by positive powers of the energy scale, and thus fade away in the infrared.

This provides justification for retaining only terms with low powers of fields. Further symmetries, such as gauge invariance, impose stringent constraints on the form of the Lagrangian \cite{Peskin1995, ChengLi1984}. Remarkably, with all its matter content and symmetries, the Standard Model involves only about twenty parameters\footnote{The number is typically quoted as 18 for the minimal model, extending to 26 if one includes CP-violating phases and right-handed neutrinos.} \cite{Branco1999, Antusch2002}, each associated with distinct symmetry-breaking patterns.

So if it is true, and no new physics enters at lower scale, the world is a ca. 20 dimensional manifold, meaning that by fixing 20 observables, the output of all other measurements is fixed, too. It sounds very strange, considering the resplendent, vibrant world of the life in Earth. But if the claim that SM describes the world is not correct, then how come the other effects? Why do not they show up in SM? According to the renormalizability arguments, if an effect, based on an operator consisting high powers of fields, plays a significant role at a given energy scale, then it must play even larger role at higher energies.

\section{The problem}

In this paper, we try to convince the reader that there is a flaw in the power counting argument. We do it in the most simple framework and speak only about electrodynamics; in particular, we will consider a system with electrons and positrons. In the standard form \cite{Peskin1995} we shall write it as
\begin{equation}
    \label{eq:plainQED}
    {\cal L} = -\frac14 F_{\mu\nu} F^{\mu\nu} +\bar\Psi_e(i \slashed D-m_e)\Psi_e+\bar\Psi_p(i \slashed D^\dagger-m_e)\Psi_p,
\end{equation}
where $\Psi_e$ and $\Psi_p$ are the electron and proton fields, and $D_\mu = \partial_\mu -ie A_\mu$ is the covariant derivative. There is no term with negative energy dimension coefficient, the masses have $d_m=2$ dimension, the gauge coupling $d_e=0$ is dimensionless.

If this form is true at a given scale, it shall remain true at lower scales, at least to describe the QED IR observables. Interactions give a logarithmic deviation from the naive power counting, in the coupling \cite{Peskin1995}
\begin{equation}
    \alpha(k) = \frac{\alpha_0}{1-\beta\alpha_0\ln k/k_0},\qquad \alpha=\frac{e^2}{4\pi},\quad \beta=\frac2{3\pi}.
\end{equation}
This formula shows that the electric coupling, although it is constant in the naive power counting, is marginally irrelevant at small scales. Using the electron mass as the potentially smallest scale, $\alpha$ reaches a constant value, which can be measured as $\sim1/137$.

For small (IR) temperatures and chemical potential, the same formulae hold. Therefore, we could conclude that at the IR, the above system consists of weakly interacting electrons and protons.

But this is far from the reality, since the electron-proton gas is not stable, bound states are formed, and in fact we will observe H-gas at low energies with a very small free electron and proton content. This recombination process is thoroughly studied, using Boltzmann equations \cite{KolbTurner1990}, and in particular this leads to the presence of microwave background radiation.

But where are the bound states in the field theory? Using MC simulations we can find them as (imaginary) poles in the ep--ep four point function, when showing an $e^{-mr}$ distance behavior. In analytic calculations we can write up an equation for the amplitude of the above scattering, where bound states show up as self-consistent poles (Bethe-Salpeter equation \cite{BetheSalpeter1951, Peskin1995}).

So we can give an account for the bound states in the field theory as static observables. But this hardly settles the matter, there are a lot of open question remaining. Let us mention some of them:
\begin{itemize}
    \item Treating bound states as static ("quenched") objects could lead to a different physics. Pure (quenched) QCD where fermions are represented as Wilson lines behave differently from the full QCD.
    \item If the bound states are dynamic, how do they show up in the renormalization group (RG) equations? If they are represented as 4-fermion operators, why do not they vanish following their power behavior?
    \item What happens if the bound state equations have several mutually excluding solutions? Which one is chosen? Are there arguments for one of them?
\end{itemize}
We will try to give the answer for all of these questions in this paper -- for the inpatient readers we suggest to skip the details, and jump directly to the Discussion section (Section \ref{sec:Discussion})!

For the more patient readers the following structure is proposed: in the next section, Section \ref{sec:scalingandrelevance}, we review once again the deep relation of scaling dimensions and relevance, and mention physical phenomena where this is applicable. In Section \ref{sec:scaling_anomailies} we will examine the fate of the 4-point fermion function in the course of RG, and try to identify the juncture that the standard reasoning overlooks. In Section \ref{sec:QED_RG} we will to be technical, and work out the relevant formulae for QED. Finally we arrive at the Discussion (Section \ref{sec:Discussion}).

\section{Scaling dimension and relevance}
\label{sec:scalingandrelevance}

There is a well established argument in the quantum (or statistical) field theories, that the relevance of a term in the Lagrangian (Hamiltonian) depends on the engineering dimension of its coupling. The argumentation goes as follows: after a proper rescaling of the physical units, there is a single scale remaining in the system, let it be an energy scale $k$. The action governing the dynamics of the system can be expanded in an operator basis, where each basis element has a definite scaling dimension. Since the action is dimensionless, each operator must have a dimensionful coupling:
\begin{equation}
    S = \sum_n g_n^{(d_n)} O_n,
\end{equation}
where $S$ is the action, $O_n$ is an operator basis, and the $d$ index indicates that the coupling $g_n^{(d_n)}$ has an engineering energy dimension $d$.

Each physical quantity can be computed from the Lagrangian, so they must be functions of the couplings. After factoring out the proper energy dimension, it is a dimensionless function. A dimensionless function, on the other hand, must only depend on dimensionless combinations. All couplings can be easily made dimensionless by rescaling with the corresponding energy scale. We find
\begin{equation}
    \label{eq:rescaledF}
    F = k^{d_F} f(  \bar g),\qquad \bar g= \{\bar g_n \,\vert\, \bar g_n=g_n^{(d_n)} k^{-d_n}\},
\end{equation}
where $F$ is our physical quantity, $d_F$ is its engineering scaling dimension, $f$ is a dimensionless function, and $\bar g$ is the set of the dimensionless rescaled versions of $g_n^{(d_n)}$.

Let us start with a generic Lagrangian defined at a high ultraviolet (UV) scale $\Lambda$, where all the rescaled couplings $\bar g_n^{(UV)}$ have order $1$ value. Then at the scale $k$ we find for the rescaled couplings
\begin{equation}
    \label{eq:engineeringRun}
    \bar g_n = g_n^{(d_n)} k^{-d_n} = \left(\frac{\Lambda}k\right)^{-d_n} \!\!\bar g_n^{(UV)}.
\end{equation}
This means that all the couplings that have \emph{negative} scaling dimension, will give contributions proportional to positive powers of $k/\Lambda$, which becomes small, if the UV scale is significantly larger that the scale of observations.

This simple observation alone gives an account for a large number of physical phenomena. Let us mention some of them:
\begin{itemize}
    \item In weak interaction, beta-decay is described by a four-fermion interaction, with interaction strength $G_F$. In four dimensions the four-fermion interaction has dimension 6, while the spacetime integration takes $4$ when we make up the action. We remain with $G_F\sim k^{-2}$. We know that at the Z-boson mass scale $m_Z$, the rescaled interaction is of order one numerical value. This means that at energy scale $E$ the strength of the weak interaction goes like
    \begin{equation}
        \bar G_F \sim \left( \frac E{m_Z}\right)^2 \bar G_F^{(UV)}.
    \end{equation}
    In the typical energy scale of the beta-decay of MeV, the ratio $E/m_Z\sim 10^{-5}$, its square is of the order of $10^{-10}$. This explains the weakness of the weak interactions, for example the extremely small cross section of the neutrino interactions.
    \item The fundamental theories of physics, the ones that make up the Standard Model, are expected to be valid over several orders of magnitude, eventually up to the unification scale $\sim 10^{10}$ GeV, or even (with slight modifications), to the Planck scale $\sim 10^{19}$ GeV. If we start from a generic system at these energies, even at the today-studied TeV scale ($10^3$ GeV) the ratio of the energy scales is at least $10^6$. Therefore all higher dimension operators can be safely neglected in our accelerator physics.
    \item The atoms consist of a rather complicated nucleus (for higher order elements), and electrons. But the size of the nucleus is about five order of magnitude smaller than the size of the atoms. This results that, apart from the conserved quantities like mass, charge and angular momentum, the complexity of the nucleus does not show up in the atomic physics.
    \item When we want to calculate the path integral with numerical methods, we must discretize the system, so, strictly speaking, we solve a different model. But the discretization effects can be described by higher dimensional operators, and this will ensure that, if the UV scale is much higher than the IR scale, their effect will be unobservable in the physical quantities.
\end{itemize}

\section{Anomalies in scaling}
\label{sec:scaling_anomailies}

We can draw the logical consequences of the previous section on QED. According to the logic described there, the Lagranigan should contain only terms with non-negative dimensional couplings, all the others are irrelevant. The relevant (renormalizable) terms are the charged fermion quadratic terms, the photon kinetic term, and their interaction. In a system consisting only of electrons and protons (for electric neutrality), the parameters of the system are $m_e$, $m_p$ for the mass of electrons and protons (having $\sim E$ dimension), and the common (dimensionless) $e$ electric charge. 

We can work out the scale dependence of the model in different approximations. In this paper we use the techniques of Functional Renormalization Group (FRG) (\cite{Wetterich1993, BergesTetradisWetterich2002, Polonyi2003, Polchinski1984}), in particular the Wetterich formalism in gauge theories \cite{Gies2004_QED, Gies2006}. We recap the computations in Section \ref{sec:QED_RG}. The main message is that below $k\approx m_e$ electron mass, the running of all couplings is stopped, and we arrive to a weakly interacting electron-proton gas.

But we know -- since observations and Boltzmann equation calculations support this claim -- that a system containing electrons and protons is not stable at low energies. The asymptotic states are not the elementary particles, but their bound states (H-atoms). The H-atom propagator, from the point of view of constituents, is a four-fermion interaction
\begin{equation}
    {\cal L}_{H-propagator} \sim \bar\psi_e \bar\psi_p \psi_e\psi_p
\end{equation}
with a proper kernel. According to our former argumentation, this is a dimension $6$ operator, meaning that its coefficient has $-2$ scaling dimension which can be neutralized by an $E^2$ factor -- exactly like in the case of the Fermi constant in the weak interaction. But what explains that in QED this operator does not die out, just in contrary, it will be the dominant degree of freedom at low energy in an electrically neutral system?

The same situation can be observed also in QCD, when the quark and gluon operators are system can consistently describe the system from infinite scale (asymptotic freedom) to zero (in appropriate RG schemes there is no IR Landau pole). But we know that the real system can be described in the language of hadrons, the bound states of quarks and gluons. Hadron propagators are four- or six quark operators, having scaling dimension 6 or 9. These seem to be irrelevant in the light of our previous argumentation, still, they are the relevant degrees of freedom of low energy QCD.

The bound state problem was addressed by several authors and papers in the literature \cite{GiesWetterich2001, BergesJungnickelWetterich1999, JakovacPatkos2019, JakovacPatkos2020}, in the context of 3D vortex confinement c.f. \cite{Fejos2020_Vortex}, in the context of Nambu-Jona-Lasinio context c.f. \cite{PhysRevD.96.076003, Braun:2018fierz, Braun:2020fierz}. In search of the answer it was proposed that in a model where both the hadronic an QCD degrees of freedom are present, as a consequence of some conspiracy in the RG equations, the RG dynamics automatically decreases the wave function renormalization of the quark and gluon operators in such an extent that it overrides the effect of the naive engineering dimension \cite{GiesWetterich2001}. However, this hope proved to be false, and there exists no such conspiracy. On the other hand the appearance and IR dominance of the bound states is a general and robust phenomenon, not just a property of the QCD. This means that an accidental conspiracy occurring in a given model can not be a good explanation for this phenomenon.

\subsection{The failure of the scaling argumentation}

Let us examine under scrutiny the arguments given above, and try to understand where it could fail.

It is certainly true that all physical quantities have a scaling dimension, which can be made dimensionless by multiplying by an appropriate factor of the scale $k$. Dimensionless physical quantities must be functions of dimensionless couplings: this also seems to be true. Therefore, \eqref{eq:rescaledF} must be valid.

The next element in the argumentation was that the most robust trend in the flow comes from the engineering dimension. Indeed, if we consider fixed points lying at weak coupling regimes -- for QED it is certainly the case -- terms containing coupling constants can not override the ${\cal O}(1)$ magnitude of the engineering dimension. Nonperturbatively, each of them will scale with an "anomalous scaling dimension", but the perturbative anomaly never reaches order one. Therefore, up to small modifications, \eqref{eq:engineeringRun} must also be valid.

On the other hand, this argument is valid only as long as the solution is regular. If singularities (poles) appear in the running of the couplings, they easily override the mild power-low behavior.

Poles are expected when an infinite number of perturbative contributions are summed up. In RG resummation this is the case, and indeed there appear poles, known as Landau poles; depending on the nature of the RG equations, they can appear in the infrared or in the ultraviolet. But also bound state formation requires the summation of an infinite number of perturbative diagrams. There, the Bethe-Salpeter equation describes that self-consistent equation whose solution yields the bound states. Also, bound states appear as poles in the higher point propagators. Therefore bound states are also naturally connected with pole-like behavior.

Technically, the Landau poles show up already in the first order RG flow. Consider the simplest quadratic flow
\begin{equation}
    \frac{d g}{d \ln k} = -\beta g^2,
\end{equation}
it has a solution
\begin{equation}
    g(k) = \dfrac1{\beta \ln \frac k {k_L}},\qquad k_L=\Lambda e^{-\frac1{\beta g(\Lambda)}}.
\end{equation}
This running exhibits singular behavior at $k=k_L$ ('L' for Landau pole). Irrespective of the value of $\beta$ (unless zero), the pole is always present.

In the appearance of Landau pole the most important factor is the unbound behavior of the momentum dependence $\ln k$. This term, on the other hand, is the consequence of the logarithmically divergent one-loop contribution to the electric charge. If this contribution is convergent, then the pole behavior disappears. To see it, let us consider a generic form
\begin{equation}
     \label{eq:matrix1loopRG}
    \frac{d g}{d \ln k} = -g \frac{d Q}{d \ln k} g,
\end{equation}
where we can choose $Q(\Lambda)=0$. Its solution is
\begin{equation}
     \label{eq:matrix-like-1loopRGsolution}
    g(k) = \dfrac{g(\Lambda)}{1+g(\Lambda) Q(k)}.
\end{equation}
If $|Q(k)|$ is bounded, let us say $|Q(k)|<Q_0$, then
\begin{equation}
     \label{eq:boundedgk}
    g(k) \le \dfrac{g(\Lambda)}{1 - g(\Lambda) Q_0},
\end{equation}
so it remains bounded, too, and no pole behavior can appear. Technically, the convergence of the loop contribution closely related with the dimension of the coupling constant, so this formula tels us again that irrelevant couplings are not necessarily to be included in the RG running.

The four-fermion interaction leads to the above type of FRG equation. Therefore, if $g$ is a scalar coupling, then we can not find a pole behavior.

But the most important property of the four-fermion interaction is that it has not a scalar coupling constant. Indeed, the coupling can depend on all four positions as well as all the four spin indices \cite{JakovacPatkos2019}. And if we interpret \eqref{eq:matrix1loopRG} as a \emph{tensor} equation, it changes the consequences dramatically.

An infinite matrix, namely, can have an unbounded spectrum even if all the matrix elements are finite. Therefore, even though $|Q_{ij}|<Q_0$ for all indices, it does not mean $|Q|<\infty$. Then \eqref{eq:boundedgk} is not true any more. In fact, the condition for having a pole is that $g(\Lambda) Q(k)$ has a $-1$ eigenvalue:
\begin{equation}
     \label{eq:matrix1loopRGpoles}
    g(\Lambda) Q(k) u(k) = -u(k).
\end{equation}
This is, identifying the roles of the symbols, exactly the Bethe-Salpeter equation \cite{BetheSalpeter1951, Peskin1995, JakovacPatkos2019}.

The fact that nonlocal structure can provide important information, was pointed out earlier by Braun and collaborators in the framework of the Nambu-Jona-Lasinio model \cite{PhysRevD.96.076003, Braun:2018fierz, Braun:2020fierz}.

This indicates that at the poles of Bethe-Salpeter equation we will observe the same pole-behavior in the corresponding RG equations like at the Landau-pole. Before we closer look at the consequences, we show that it indeed happens in QED. In the following section technical details come, the reader who does not care for them, can safely drop to the Discussion part (Section \ref{sec:Discussion}).

\section{The RG equations of the extended QED}
\label{sec:QED_RG}

Let us examine the extended form of the U(1) gauge theory containing, besides quadratic electron and proton terms, their four point function, where we expect the H-atom-like poles show up:
\begin{equation}
    \label{eq:extendedQED}
    {\cal L} = -\frac14 F_{\mu\nu} F^{\mu\nu} +\bar\Psi_e(i \slashed D-m_e)\Psi_e+\bar\Psi_p(i \slashed D^\dagger-m_e)\Psi_p + \lambda \Psi_e^\dagger \Psi_p^\dagger\Psi_e\Psi_p,
\end{equation}
 and $\lambda$ is a general tensor that depends on the momentum and spin indices of the four fermions it is attached to:
\begin{equation}
    \lambda \Psi_e^\dagger \Psi_p^\dagger\Psi_e\Psi_p \equiv \lambda_{\alpha\beta\gamma\delta}(p,q,k,\ell) \Psi_{e\alpha}^\dagger(p)\Psi_{p\beta}^\dagger(q)\Psi_{e\gamma}(k)\Psi_{p\delta}(\ell)(2\pi)^4 \delta(p+q-k-\ell).
\end{equation}
We will simply ask whether the four fermion term is relevant or not from the point of view of the RG flow.

Let us write up formally the FRG equations of the extended QED \eqref{eq:extendedQED}. These can also be found in \cite{JakovacPatkos2019}, here we repeat the calculations in a somewhat different point of view. 

First, we recap the treatment of the RG flow without the extra term, and in the next subsection we write up the running of the 4-point function. We introduce the following renormalized Lagrangian
\begin{align}
    \label{eq:RGQED}
    {\cal L} =& -\frac {Z_3}4 F_{\mu\nu} F^{\mu\nu} 
    + Z_2^{(e)} \bar\Psi_e i \slashed \partial \Psi_e - m_e \bar\Psi_e \Psi_e + Z_1^{(e)} e \bar\Psi_e \slashed A \Psi_e +\cr
    & +Z_2^{(p)} \bar\Psi_p i \slashed \partial \Psi_p - m_p \bar\Psi_p \Psi_p - Z_1^{(p)} e \bar\Psi_p \slashed A \Psi_p.\cr
\end{align}
The formulae for the different contributions can be derive from the one loop self energies and the one loop charge renormalization. The calculations can be found in Appendix \ref{sec:FRGbetacalc}.

Ward identities dictate $Z_1=Z_2$ for both electron and proton. This means that the Ansatz we use may look like
\begin{align}
    {\cal L} =& -\frac {Z_3}4 F_{\mu\nu} F^{\mu\nu} 
    + Z_2^{(e)} \bar\Psi_e (i \slashed \partial + e \slashed A) \Psi_e - m_e \bar\Psi_e \Psi_e\cr
    & +Z_2^{(p)} \bar\Psi_p (i \slashed \partial - e\slashed A) \Psi_p - m_p \bar\Psi_p \Psi_p.\cr
\end{align}
This means that there are five renormalization constants to evolve, where the running of the proton and electron renormalization constants are analogous. We find
\begin{align}
    \frac{d\ln Z_3}{d\ln k} &= -\frac{e^2Z_3^{-1}}{6\pi^2} \left[\frac1{(1+M_e^2/k^2)^3} + \frac1{(1+M_p^2/k^2)^3}\right],\cr
    \frac{d \ln M }{d\ln k} &= \frac{e^2 Z_3^{-1}}{2\pi^2} \frac1{(1+M^2/k^2)^2},\cr
    \frac{d\ln Z_2}{d\ln k} &= -\frac{e^2Z_3^{-1}}{8\pi^2} \frac1{(1+M^2/k^2)^2},\cr
\end{align}
where $M=m/Z_2$. With the usual $\alpha$ notation
\begin{equation}
    \alpha = \frac{e^2}{4\pi Z_3}
\end{equation}
we find
\begin{align}
    \label{eq:pureQED}
    \frac{d\alpha}{d\ln k} &= \frac{2\alpha^2}{3\pi} \left[\frac1{(1+M_e^2/k^2)^3} + \frac1{(1+M_p^2/k^2)^3}\right],\cr
    \frac{d \ln M }{d\ln k} &= \frac{2\alpha}{\pi} \frac1{(1+M^2/k^2)^2},\cr
    \frac{d\ln Z_2}{d\ln k} &= -\frac{\alpha}{2\pi} \frac1{(1+M^2/k^2)^2}.\cr
\end{align}
As we see, the $Z_2$ decouples from the evolution, it just follows the scaling of the others.

For general scale, the equations are not independent, and are only numerically solvable. If $k\gg M$, then we can omit the $M^2/k^2$ terms, and we obtain
\begin{equation}
    \alpha(k \gg M_p) = \dfrac{\alpha_0}{1-\frac{2N_f}{3\pi}\alpha_0 \ln\frac k{k_0}},\qquad M=M_0\left(\frac{\alpha(k)}{\alpha(k_0)}\right)^3,
\end{equation}
where $N_f$ is the number of fermion species, now it is 2. In the other limit $k\ll M_e$, and the solution reads:
\begin{equation}
    \alpha(k\ll M_e) = \alpha(k=0) + \frac{2k^6}{18\pi M_e^6}\alpha^2(k=0),\qquad M_e^4 = M_{e0}^4 + \frac{2\alpha(k=0)}{\pi} k^4,
\end{equation}
where $\alpha(k=0)\approx1/137$. As it can be seen, below $k\approx M$ the running of all couplings is stopped.

\subsection{The electron-proton scattering interaction}

Now let us turn to the FRG evolution equation of the electron-proton scattering term. At one loop level the four-fermion term does not influence the running of the coupling. Therefore in the FRG calculations the running of the $Z$ factors govern the running of $\lambda$, but there is no backreaction. The case is similar to the running of $Z_2$: this is governed by the masses and $Z_3$, but there is no backreaction.

The detailed calculation can be found in \cite{JakovacPatkos2019}. In the running of $\lambda$ there are explicit terms of order $e^4$ and self-consistent terms. 

\subsubsection{Explicit contribution to the running of $\lambda$}
If we considered only the explicit terms, we would have
\begin{align}
    \label{eq:4pointfunctionrun}
     \partial_k\lambda_{\alpha\beta\sigma\delta}(p) = &e^4 \partial'_k \int\frac{d^4\ell}{(2\pi)^4} \frac1{\ell^2}\frac1
    {(\ell+p_1-p_2)^2} \left[\gamma^\mu \frac1{\slashed p_1 + \slashed \ell - m_e} \gamma^\nu\right]^{(reg)}_{\beta\alpha}\times\cr
     & \times\biggl\{ \left[\gamma_\mu \frac1{\slashed p_3 -\slashed \ell - m_p} \gamma_\nu\right]^{(reg)}_{\delta\sigma} + \left[\gamma_\nu \frac1{\slashed p_4 + \slashed \ell - m_p} \gamma_\mu\right]^{(reg)}_{\delta\sigma} \biggr\},
\end{align}
where $p=(p_1,p_2,p_3,p_4)$ composite momentum variable and the regulated fermion propagator is
\begin{equation}
    \left[\frac1{\slashed p-m}\right]^{(reg)} = \frac1{ \left(\Theta(p-k)+\dfrac kp\Theta(k-p)\right) \slashed p - m},
\end{equation}
and the $\partial'$ operator acts only on the regulator.

We do not want to calculate this term explicitly, just concentrate on some of its general properties. First of all, if we neglect the $k$ dependence of $e$ an the fermion masses, then the above formula can be explicitly integrated out. This expression is UV finite: for large internal momenta its contribution reads
\begin{equation}
    e^4 \int\frac{d^4\ell}{(2\pi)^4} \frac{(\gamma^\mu\slashed\ell \gamma^\nu)_{\beta\alpha} [(\gamma_\mu\slashed\ell \gamma_\nu)_{\delta\sigma} - (\gamma_\nu\slashed\ell \gamma_\mu)_{\delta\sigma}]}{\ell^8},
\end{equation}
which is finite.

To estimate the approximate $k$ dependence we may argue that the two regulated fermion propagators will yield $\sim1/(k^2+M^2)$ term, where $M$ is some mass scale composed of the external momenta and fermion masses. The $1/k^2$ dependence fits to the engineering dimension of $\lambda$ which is $-2$. Therefore we expect that the running of $\lambda$ behaves as
\begin{equation}
    \partial_k\lambda(k) \sim \frac{e(k)^4 k}{(k^2+M^2)^2}.
\end{equation}
For large scales the value of $\lambda$ hardly changes, for small $k$ it picks up a correction of order $e^4$. So we expect
\begin{equation}
    \frac{\lambda(k)}{\lambda(\Lambda)} = 1+ \frac{Ce(k)^4}{k^2+M^2},
\end{equation}
where $C$ is some coefficient. This means for the rescaled couplings:
\begin{equation}
    \frac{\bar\lambda(k)}{\bar\lambda(\Lambda)} = \frac{k^2}{\Lambda^2}\left(1 + \frac{Ce(k)^4}{k^2+M^2}\right).
\end{equation}
Starting with a rescaled coupling of order unity, the rescaled coupling $\bar\lambda(k)$ goes as $\sim k^2/\Lambda^2$. As it was expected, the explicit contributions render this term irrelevant in the IR.

\subsubsection{Self-consistent $\lambda$ evolution}

Let us now take into account the back-reaction of the coupling $\lambda$ to its own scale evolution. In order to get rid of long formulae, we make a simplification in the notation. We will use electron-proton pairs as states, their state is characterized by the multi-index $p\alpha\delta$ and an additional index $\ell$. Here $p$ is the electron four-momentum, $\alpha$ and $\delta$ are the Fermi indices of the electron and proton, respectively, and $\ell$ is the total momentum of the e-p system (therefore the momentum of the proton is $p'=\ell-p$). Note that in the $s$-channel processes $\ell$ is conserved, and so it is a spectator index of the diagrams.

The four-fermion coupling we are looking after can be written as
\begin{equation}
    \lambda_{\alpha\beta\sigma\rho}(p,q,\ell-p,\ell-q) = {\bm \lambda}^{(\ell)}_{p\alpha\sigma,q\beta\rho}.
\end{equation}
In this multi-index form, $\bm\lambda$ is a self-adjoint matrix.

In the multi-index notation the one-photon exchange can be represented as a matrix
\begin{equation}
    \label{eq:Vdef}
    {\bm V}_{p\alpha\sigma, q\beta\rho} = \frac{e^2\gamma^\mu_{\beta\alpha} \gamma_{\mu\rho\sigma}}{(p-q)^2+i\varepsilon}.
\end{equation}
This term does not depend on $\ell$. The other ingredient is the 2-fermion propagator, again a matrix in the multi-index space:
\begin{equation}
    {\cal G}^{(\ell)}_{p\alpha\sigma,q\beta\rho} = G^{(e)}_{\beta\alpha}(p) G^{(p)}_{\sigma\rho}(\ell-p) \delta_{pq}. 
\end{equation}
This depends on the spectator index, but it is diagonal in the momentum indices.

With this notation the $s$-channel part of the right hand side of \eqref{eq:4pointfunctionrun} can be written as a matrix multiplication
\begin{equation}
    {\bm V} {\cal G}^{(\ell)} {\bm V}.
\end{equation}

We obtain similar contributions from the diagrams where instead of ${\bm V}$ we substitute ${\bm \lambda}$. Their total $s$-channel contributions reads
\begin{equation}
    (\bm V + \bm \lambda) {\cal G}^{(\ell)}(\bm V + \bm \lambda).
\end{equation}

The FRG equation according to this formula reads
\begin{equation}
    \partial_k \bm \lambda = \partial'_k (\bm V + \bm \lambda) {\cal G}^{(\ell)}(\bm V + \bm \lambda) + \mathrm{t-channel}.
\end{equation}
In the followings we omit the t-channel contribution, since it does not contribute to the pole behavior \cite{Nakanishi1969}.

Now let us assume that we apply an FRG regulator that only restricts the fermion momenta \cite{JakovacPatkos2019}. In this case the explicit $k$-dependence comes from ${\cal G}$, and so
\begin{equation}
    \partial_k \bm \lambda = (\bm V + \bm \lambda) \partial_k{\cal G}^{(\ell)}(\bm V + \bm \lambda).
\end{equation}
If $\bm\lambda$ is small, we can observe that behaviour that was analysed before. At $k=0$ practically the one loop formula is reproduced
\begin{equation}
    \bm\lambda(k=0) \approx \bm V{\cal G}^{(\ell)}\bm V.
\end{equation}
In the $\lambda>0$ case we expect some modifications. To be able to keep track of this change, we assume that we arrived with the scale below $m_e$. There all the QED couplings have constant values, and then $\bm V$ does not depend on $k$. Therefore we can introduce
\begin{equation}
    \bm{\bar \lambda} = \bm V + \bm \lambda,
\end{equation}
for which we can write
\begin{equation}
    \label{sec:lambdaRG}
    \partial_k \bm{\bar \lambda} = \bm{\bar \lambda} \partial_k{\cal G}^{(\ell)}\bm{\bar \lambda}.
\end{equation}
This equation can be solved
\begin{equation}
    {\bm{\bar \lambda}}_k = (1+ {\bm{\bar \lambda}}_\Lambda {\cal G})^{-1} {\bm{\bar \lambda}}_\Lambda.
\end{equation}
The value of  ${\bm{\bar \lambda}}_\Lambda$ with cutoff value around the electron mass is almost $\bm V$, since the $\bm\lambda$ part gives an ${\cal O}(e^4)$ correction to it. Therefore we can write
\begin{equation}
    \label{eq:lambdasolution}
    \bm\lambda_k = (1+ \bm V {\cal G}^{(\ell)})^{-1}\bm V{\cal G}^{(\ell)}\bm V.
\end{equation}
These equations are exactly those that were listed in equations \eqref{eq:matrix1loopRG}-\eqref{eq:matrix1loopRGpoles}. This means that the scenario sketched there indeed manifests itself in the extended QED.

\section{Discussion: bound state pole in the renormalization group}
\label{sec:Discussion}

The formula we obtained, eq. \eqref{eq:lambdasolution}, actually, is not surprising at all. The summation of the s-channel diagrams is analogous to the propagator self-energy resummation, so we expect a formula that looks like the propagator dressed with the self energy. And indeed, \eqref{eq:lambdasolution} is of this form.

Now we can identify those statements we assumed earlier about the structure of the FRG running of the four-fermion term. In the formula the higher loop corrections are suppressed by $\alpha$ (c.f. the definition of $\bm{V}$ in \eqref{eq:Vdef}). Thus, naively, we do not expect that the "denominator" $(1+ \bm V {\cal G}^{(\ell)})$ can be zero, since ${\cal G}$ is order $1$ and $V$ is order $\alpha$, their product may never reach the order of 1. Still, the equation describing the appearance of the pole
\begin{equation}
    \label{eq:lambdapole}
    \bm V {\cal G}^{(\ell)} \bm u_0 = -\bm u_0
\end{equation}
is \emph{known} to have solution in the IR. This seems to be a contradiction, which has to be understood.

As we discussed earlier, the solution lies in the matrix nature. Here we deal with infinitely large matrices, with unbounded spectrum. The factor that compensates the $\alpha$ suppression can be the number of matrix elements involved in the eigenvectors. We expect that summing $N\sim 1/\alpha$ number of rows can overcome the $\alpha$-suppression. Since $\alpha$ is small, this means a lot of modes. Actually, this $1/\alpha$ dependence can be associated with the fact that the H-atom has a finite size proportional to this factor.

Considering scale dependence, in the UV everything is regular, while in the IR we know that \eqref{eq:lambdapole} has a lot of solutions. So, in the course of the RG flow there is a scale, where the first pole appears.

To see the pole behavior, we rewrite the \eqref{eq:lambdasolution} somewhat, to make the self-adjoint nature explicit:
\begin{equation}
    \lambda_k = \sqrt{\bm V}\, \frac{\sqrt{\bm V}{\cal G}^{(\ell)}\sqrt{\bm V}}{1+\sqrt{\bm V}{\cal G}^{(\ell)}\sqrt{\bm V}}\,\sqrt{\bm V}.
\end{equation}
This is a correct form, since in the fraction the numerator and the denominator consists of the same matrix.

Now let us assume that at the scale $K^{(\ell)}$ the denominator vanishes (note that it is $\ell$-dependent). We separate the singular part as
\begin{equation}
    \sqrt{\bm V}{\cal G}^{(\ell)}\sqrt{\bm V} = (-1+c^{(\ell)}(k-K^{(\ell)}))\Pi_0^{(\ell)} + Q^{(\ell)}
\end{equation}
where $\Pi_0^{(\ell)} = \bm u_0^{(\ell)} \otimes {\bm u_0^{(\ell)}}^\dagger$ is the projector on the eigenspace where the $-1$ eigenvalue shows up, and $Q^{(\ell)}$ denotes the regular part. Then the four point function reads
\begin{equation}
    \lambda_{k\approx K^{(\ell)}} = \frac {{\bm v^{(\ell)}}\otimes {\bm v^{(\ell)}}^\dagger}{k-K^{(\ell)}}  + \mathrm{regular\ part},\qquad \bm v^{(\ell)} = \sqrt{c^{(\ell)}\bm V} \bm u_0^{(\ell)}.
\end{equation}

This pole behavior exceeds all suppression when we approach $k=k_0$ scale. Therefore neglecting the 4-fermion interaction is only justified \emph{above} the $k_0$ scale. Below this scale the usual QED RG flow equations are no longer valid.

To be able to go on with the lowering of the scale, this singularity should be removed. We can use the regular four-point function
\begin{equation}
    \lambda^{reg}_k = \lambda_k - \sum_\ell \frac{\bm v^{(\ell)}\otimes {\bm v^{(\ell)}}^\dagger}{k-K^{(\ell)}},
\end{equation}
The singularity then is isolated in the Lagrangian as
\begin{equation}
    \Psi_e^\dagger \Psi_p^\dagger \lambda^{singular} \Psi_e\Psi_p = \sum_\ell \frac1{k-K^{(\ell)}}|{\bm v^{(\ell)}}^\dagger \Psi_e\Psi_p|^2.
\end{equation}
We can represent this term using the Hubbard-Stratonovich transformation, and find
\begin{align}
    \label{eq:QEDandH}
    {\cal L} &= -\frac14 F_{\mu\nu} F^{\mu\nu} +\bar\Psi_e(i \slashed D-m_e)\Psi_e+\bar\Psi_p(i \slashed D^\dagger-m_e)\Psi_p + \lambda^{reg} \Psi_e^\dagger \Psi_p^\dagger\Psi_e\Psi_p + \cr
    &+ \sum_\ell (\frac12 {H^{(\ell)}}^\dagger (K^{(\ell)}-k) H^{(\ell)} + H^{(\ell)} \bm v^{(\ell)} \Psi_e\Psi_p + \mathrm{h.c.}) .
\end{align}
This form is equivalent to the extended QED \eqref{eq:extendedQED} above $k_0$, but it can also be applied below the would-be poles of the original Lagrangian. The pole behavior of $\lambda$ transfers to the pole behavior of the H-propagator, thus rendering it a real, long-living particle at low scales.

With this construction we could get rid of the first poles, but, of course, further poles are following this at lower scales. At $k=0$ IR scale we have an infinitely many poles. To overcome the corresponding singularities, we have to introduce new and new degrees of freedom, represented as real particles, whose pole in their propagator exactly cancels the pole in the 4-point function. Finally, we arrive at a Lagrangian with infinitely many H-degrees of freedom. The regular part can be kept, as we did in the above formulae, or fully represented by scattering modes, as in \cite{JakovacPatkos2019}.

We have to emphasize also that fact that, although the singular behavior is associated to the new term $H\bm v \Psi_e\Psi_p$, but this term it alone leads a regular RG. The build-up of the singularity requires the presence of all the modes in the 4-point function.

\section{Conclusion}
\label{sec:Conclusion}

In this paper we examined, what happens, if we extend QED with a non-renormalizable 4-point interaction term, starting its coupling with an order one value in the far UV.

At high scales, as expected, the rescaled 4-point function decreases with it engineering dimension 2 (omitting anomalous dimension, which is small in this case). Far from the UV starting point, therefore, its value is negligible. This is the normal RG flow of an irrelevant operator.

\begin{figure}
    \centering
    \includegraphics[width=0.5\linewidth]{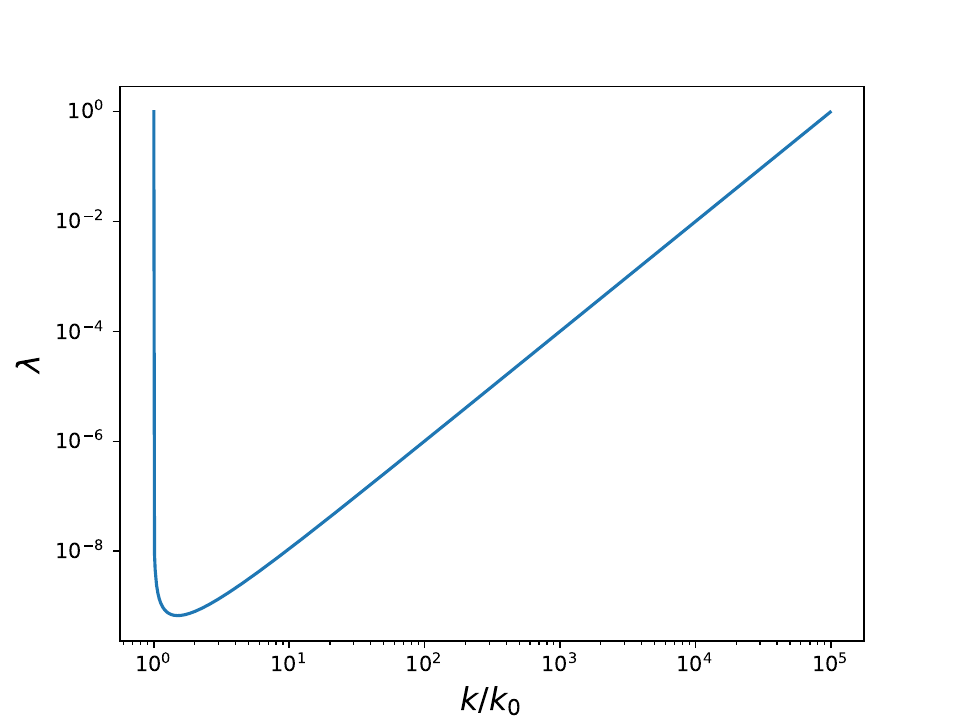}
    \caption{The qualitative behavior of the rescaled 4-fermion coupling. At high scale, even if it starts with a value around unity, it decreases with a power law, rendering it irrelevant at lower scales. At a certain $k=K$ scale, however, it reaches a pole, and it becomes relevant again.}
    \label{fig:lambdarun}
\end{figure}

But at lower scales something unexpected happens, because the running coupling reaches a pole at a certain scale. Since $f(k)/(k-K)\to\infty$ (provided $f(K)\neq0$), the rescaled four-point function becomes relevant, despite the fact that it was negligible at somewhat higher scale. The qualitative running of the four-point function is shown in Fig.~\ref{fig:lambdarun}.

This means that the QED, in the renormalizable form, as we know it from textbooks, can not give an account for the IR behavior, as the RG flow is considered. 

This observation naturally raises a lot of questions. The most important is that if we can not determine by simple methods which operators become relevant at lower scale, what Lagrangian should we use? And the sad answer is that we do not know. Any multiparticle operators can be singular in certain circumstances, or at least it can not be said from the UV whether it becomes singular anywhere or not. The only way to know this is to perform the RG scaling, and check whether it stays irrelevant in the IR, or not.

The next question is that numerical methods based on a given form of the theory can give an account for the correct IR behavior or not. And the answer here is, again, not fully supportive.

The point is that if we did not restricted ourselves to the UV form of the QED Lagrangian in the FRG running, then the full effective action at a certain scale would have contained a lot of terms, most of them are irrelevantly small. But still, the act as a seed for an exponential growth at lower scale. Thus, the \emph{exact} computations in QED, even when we started from the purely renormalizable form at a given UV scale, lead to the appearance of H-atoms (or other bound states) at lower scales.

But there is a caveat. If, namely, there are \emph{several}, mutually exclusive fixed points in the RG flow, that are separated from each other by separatrices, then it is not the same, which UV action do we use, even if they are the same up to perturbatively irrelevant terms.

So, it is absolutely conceivable that we start from two different, but equivalent actions, compute the IR physics, and we get different results. To be profane, using one form of the SM yields a stone in the IR, using another one, we obtain a cake. A stone and a cake are the same in the UV, but at a certain scale they start to become distinguishable. And those operators that make the distinction between them, are all irrelevant from the point of view of the SM.

This means that the dream that knowing the SM is equivalent to knowing the whole world is not true. Even with the smartest simulation we can not find all the phenomena occurring in IR, since they are determined by those parts that are irrelevant in the UV.

This can be considered as a bad news or a good news, depending on our approach. It is bad in the sense that the predictive power of our UV models are limited, actually limited to their own relevant scales. The good news is that the world is rich, infinitely rich in structures, allows all kind of unexpected phenomena, and there is no logical microscopic reason that certain constructions in IR shall not be realized.

\section*{Acknowledgment} the author acknowledges discussion with A. Patkós, G. Fejős and T.S Bíró that contributed to the understanding of the topic.

\bibliographystyle{unsrt} 
\bibliography{renormalizability_references}

\section*{Appendix}

\appendix
\section{FRG calculation of the beta functions}
\label{sec:FRGbetacalc}

The individual propagators are:
\begin{align}
    G^{(fermion)}(p) &= \frac1{Z_2 \slashed p + m} = \frac{Z_2^{-1}}{\slashed p +M},\qquad M=\frac{m}{Z_2}\cr
    G^{(A)}_{\mu\nu}(p) &= \frac{g_{\mu\nu}Z_3^{-1}}{p^2}\cr
\end{align}
and the vertex is
\begin{equation}
    Z_1 e \gamma^\mu.
\end{equation}
From these ingredients we calculate the self energies and the radiative correction to the charge.

The photon self energy contribution reads (Peskin Schroeder, p.246) from a single fermion species
\begin{align}
    i\delta\Pi_2^{\mu\nu}(q) = -4C_A \int \frac{d^4k}{(2\pi)^4} \Tr \frac{k^\mu(k+q)^\nu + k^\nu(k+q)^\mu - g^{\mu\nu}(k(k+q)-M^2}{(k^2-M^2)((k+q)^2-M^2)},
\end{align}
where
\begin{equation}
    C_A=Z_1^2 Z_2^{-2}e^2.
\end{equation}
With Feynman parametrization, after Wick rotation we find
\begin{align}
    \delta\Pi_2^{\mu\nu}(q) = -4C_A\int\limits_0^1 dx \int\frac{d^4\ell}{(2\pi)^4} \frac{g^{\mu\nu}(\ell^2/2+M^2+q^2x(1-x)) -2x(1-x)q^\mu q^\nu} {(\ell^2 +M^2-x(1-x)q^2)^2}.
\end{align}
We expect
\begin{align}
    \delta\Pi_2^{\mu\nu}(q^2) = \delta\Pi_2(q^2)(q^2g^{\mu\nu} - q^\mu q^\nu).
\end{align}
We need $\Pi_2(0)$ which reads
\begin{align}
    \delta\Pi_2(0) = -8C_A \int\limits_0^1 dx \int\frac{d^4\ell}{(2\pi)^4} \frac{x(1-x)} {(\ell^2 +M^2)^2} = -\frac{4e^2Z_1^2 Z_2^{-2}}3 \int\frac{d^4\ell}{(2\pi)^4} \frac1{(\ell^2 +M^2)^2}.
\end{align}
We can safely perform the angle integral
\begin{align}
    \delta\Pi_2(0) =-\frac{C_A}{6\pi^2} \int d\ell \frac{\ell^3}{(\ell^2 +M^2)^2}.
\end{align}
In cutoff regularization it yields the usual $\log\Lambda$ divergence. In the FRG approach we regularize the fermion momenta and write
\begin{align}
    \delta\Pi_{2k}(0) =-\frac{C_A}{6\pi^2} \int d\ell \frac{\ell^3}{(\ell^2\Theta(\ell-k)+k^2\Theta(k-\ell) +M^2)^2}.
\end{align}
Its derivative with respect to $k$ reads
\begin{align}
    \partial_k \delta\Pi_{2k}(0) = \frac{C_A}{6\pi^2} \int d\ell \frac{4k \Theta(k-\ell)\ell^3}{(\ell^2\Theta(\ell-k)+k^2\Theta(k-\ell) +M^2)^3},
\end{align}
which finally leads to
\begin{equation}
     \partial_k \delta\Pi_{2k}(0) = \frac{e^2Z_1^2 Z_2^{-2}}{6\pi^2} \frac{k^5}{(k^2+M^2)^3}.
\end{equation}

In a similar manner the fermion self energy reads (c.f. Peskin-Schroeder p217. \cite{Peskin1995}):
\begin{align}
    \Sigma(p) = -iC_\Psi \int \frac{d^4q}{(2\pi)^4} \gamma_\mu \frac{\slashed q+m}{q^2-M^2} \gamma^\mu \frac1{(p-q)^2}.
\end{align}
where
\begin{equation}
    C_\Psi =e^2Z_1^2Z_3^{-1}Z_2^{-1}.
\end{equation}
After performing the Clifford algebra manipulations we find
\begin{align}
    \Sigma(p) = -iC_\Psi \int \frac{d^4q}{(2\pi)^4} \frac{-2\slashed q+4M}{q^2-M^2}\frac1{(p-q)^2}
\end{align}
Expanding around zero we find
\begin{equation}
    M\Sigma_0 + \slashed p \Sigma_1 = (-iC_\Psi)(4M-\slashed p)\int \frac{d^4q}{(2\pi)^4} \frac1{q^2-M^2}\frac1{q^2}.
\end{equation}
Performing Wick rotation, and then doing the angular integration results in
\begin{equation}
    \frac14\Sigma_0= -\Sigma_1 = \frac{C_\Psi}{8\pi^2} \int dq \frac q{q^2+M^2}.
\end{equation}
In the FRG formalism only the fermion momenta are regularized, which results is
\begin{equation}
    \partial_k \Sigma_1 = \frac{e^2Z_1^2Z_3^{-1}Z_2^{-1}}{8\pi^2} \frac{k^3}{(k^2+M^2)^2}.
\end{equation}

For the charge renormalization it is enough to consider the one loop diagram at zero external momenta. For one fermion species we have (Peskin-Schroeder p189)
\begin{align}
    \Gamma^\mu = iC_\Gamma \int\frac{d^4q}{(2\pi)^4} \frac1 q^2 \gamma^\nu \frac1{\slashed q - M} \gamma^\mu \frac1{\slashed q - M} \gamma_\nu,
\end{align}
where
\begin{equation}
    C_\Gamma =e^3 Z_1^3 Z_2^{-2} Z_3^{-1}.
\end{equation}
After expanding the gamma algebra we find
\begin{align}
    \Gamma^\mu = iC_\Gamma \gamma^\mu \int\frac{d^4q}{(2\pi)^4} \frac{-q^2 + 2M^2}{q^2(q^2-M^2)^2}.
\end{align}
After performing the angular integration and performing Wick rotation we find
\begin{align}
    \Gamma^\mu = \frac{C_\Gamma}{8\pi^2} \gamma^\mu \int dq q\frac{q^2 + 2M^2}{(q^2+M^2)^2}.
\end{align}
In the UV it yields the same UV divergence than that of coming from $\Sigma(p)$. In FRG we find
\begin{equation}
    \partial_k \Gamma^\mu = \frac{e^3 Z_1^3 Z_2^{-2} Z_3^{-1}}{8\pi^2}\gamma^\mu \frac{k^3}{(k^2+M^2)^2}\left(1+\frac{2M^2}{k^2+M^2}\right).
\end{equation}

The Ward identity require
\begin{align}
    q_\mu \Gamma^\mu(0) -G^{-1}(p+q)+G^{-1}(p) \stackrel{q\to0}\longrightarrow 0.
\end{align}
This requires $Z_1=Z_2$, for all fermion species. We obtain a slight violation of the Ward identities here. Putting the requirement of charge conservation first \cite{Gies2004_QED}, we use the simpler formula for $Z_1$, too.

For small external momenta $p_i\to 0$, denoting the momentum transfer by $Q=p_2-p_1$, we may omit all momentum dependence except the photon $(\ell-Q)^2$. We find for the magnitude
\begin{equation}
    C_1(p\approx0) \sim e^4 \int \frac{d^4\ell}{\ell^2(\ell-Q)^2} \sim e^4 \ln \frac{m_e}Q,
\end{equation}
where the numerator can be another mass scale. It has a logarithmic divergence for small momentum transfer, which is the signal of the soft divergences of QED. As a coefficient of a local operator, however, we shall integrate over the external momenta, and so the logarithmic divergence does not cause problem in the magnitude.

\end{document}